# Système d'indexation et de recherche de vidéo intégrant un système gestuel pour les personnes handicapées

Mohamed HAMROUN, Mohamed Salim BOUHLEL

*UR SETIT, ISBS Université de Sfax - Tunisie*

**hamroun.mohammed@gmail.com**
**medsalim.bouhlel@enis.rnu.tn**

**Résumé** : La quantité d'informations audiovisuelles s'est accrue de façon spectaculaire avec l'apparition de l'Internet à haut débit et de la télévision numérique. De plus, les avancées technologiques réalisées ces dernières années dans le domaine de l'informatique (espaces de stockage de plus en plus considérables, numérisation des données, etc.) ont permis de simplifier l'utilisation de données vidéos dans divers domaines par le grand public. Ce qui a rendu possible le stockage des grandes collections de documents vidéo dans des systèmes informatiques. Pour permettre une exploitation efficace de ces collections, il est nécessaire de mettre en place en place des outils facilitant l'accès à leurs documents et la manipulation de ceux-ci. Dans cet article nous proposons une méthode de recherche multimédia dans une base de données vidéo, en se basant sur une technique de pondération pour dégager le degré d'appartenance d'un concept dans une vidéo aussi que sur la structuration de données de la base audio-visuelle (contexte/concept), Ainsi que sur un mécanisme de retour de pertinence. Enfin, Nous avons décidé de créer un système de recherche, offrant en plus des commandes usuelles, différents types d'accès au système, en fonction du handicap de la personne. En effet, l'application se compose d'un système de recherche, mais offre l'accès aux commandes par le biais de la voix ou les gestes.
**Mots clés :** Document vidéo, moteur de recherche, concept, contexte, réinjection de pertinence, requête textuelle, requête image, navigation, système gestuel, système vocale, kinect.

## 1. Introduction

Les données multimédias (texte, image, vidéo et son) contiennent une densité variable d'informations et beaucoup de redondances. Un utilisateur cherchant un média s'attend à trouver des informations sémantiques alors qu'on ne peut, dans la plupart des cas, n'extraire directement que des données structurelles. En plus, l'utilisateur ne sait pas toujours exprimer sa requête dans un langage formel. On peut par exemple chercher des images « fleurs », dans ce cas il s'agit d'images sur fond vert ayant la texture connue de l'herbe avec un objet circulaire de couleur plus ou moins vive. Il peut aussi chercher des photographies de chevaux, dans ce cas un cheval alezan (de robe brune) dans une prairie sera aussi pertinent qu'un cheval blanc dans une carrière. Cette fois les couleurs ne sont plus discriminantes, seule la forme permet de trouver les images souhaitées. Ainsi deux requêtes similaires sémantiquement seront très différentes formellement. Il est donc important de définir une interface intuitive permettant à l'utilisateur d'atteindre l'ensemble des objets qu'il recherche.

Pour ces raisons la visualisation de l'information est devenue une zone de recherche très important. La recherche graphique figure comme une alternative aux procédés de recherche impliquant à l'utilisateur la charge de bien spécifier leurs requêtes pour débuter un processus d'exploration des bases audiovisuelles. Rechercher par simples navigation ou avec une requête image requière cependant de l'utilisateur une connaissance a priori des données ainsi que de leur structure.

Un processus graphique, exige, cependant, une organisation a priori de la collection des données ainsi que l'intégration d'outils permettant d'optimiser l'interaction entre le navigateur et la collection.
Notre contribution vise une finalité similaire. Nous nous proposons d'intégrer un contexte de recherche graphique basé sur une organisation sémantique de données qui a pour intérêt de procurer à l'utilisateur une représentation des connaissances proche du modèle cognitif. Il s'agit également d'aider l'utilisateur à retrouver le plus rapidement possible les connaissances qu'il cherche en lui offrant des moyens de représentation et de visualisation lui permettant de filtrer les connaissances et sélectionner uniquement celles qui sont pertinentes pour lui.

Plusieurs travaux proposent une méthode de recherche basée sur cette notion, nous avons dressé le tableau 1 qui offre une vue synthétique et résumée des propositions référencées, sous plusieurs points de vue : les éléments de navigation, le corpus de vidéos et les techniques de visualisation proposée ainsi que certains systèmes supportent deux modes de recherche (recherche par navigation et par requêtes).

| Système de recherche | Eléments de Navigation | Corpus de vidéo | Technique de visualisation |
|---|---|---|---|
| B"arecke et al (VidéoSOM) [5] | plan | Vidéo de type News issue de Collection TRECVID 2006 | Représentation temporelle de plan, |
| Goeau et al (TGM) [6] | Plan | News, émissions sportives | Graphe 2D, technique de zoom géométrique |
| Rooij et al.et Snoek et al [7,8] | Plan | TRECVID 2007 | navigation visuelle basée sur les threads |
| Heesch et al [9] | images –clés | TRECVID2004 | Storyboard, graphe |
| Anthony Don [10] | Images- clés | TRECVID 2002 | Graphe 2D, technique de zoom géométrique |

Tableau 1 : Etude comparative des systèmes existants

Nous visons par ce travail de proposer une méthode de recherche vidéo qui met l'utilisateur au centre d'intérêt. L'intervention de ce dernier dans le processus de recherche nous dévoile ses besoins et ses désirs. Notre amélioration se manifeste, premièrement, dans la façon de regrouper et d'organiser les données pour faciliter l'exploitation de données lors de la recherche :

Nous avons pris, comme point de départ, une base vidéo indexée à base de concepts [1].

## 2. Description de notre SRI

Dans notre première contribution, nous avons proposé une méthode du calcul de la similarité inter-concepts basée sur deux sources d'information : le corpus audio-visuel et l'ontologie à fin de mettre en place des focus (1) qui présentent un point de départ pertinent pour le processus de recherche.

Nous avons intéressé également à la structuration du corpus audio-visuel l (2) pour pallier le problème de la quantité et l'hétérogénéité du contenu

La deuxième contribution consiste à proposer des techniques de visualisation interactives (3) qui reposent sur un fort couplage entre les vues globales et les vues locales pour faciliter l'accès à la collection

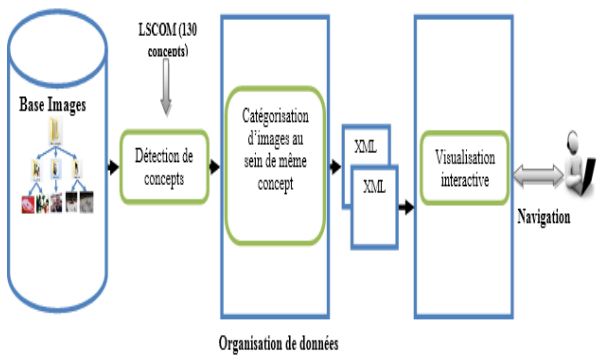

Figure 1. Architecture conceptuelle de système proposé

### 2.1. Mesure de similarité inter-concept proposée

Nous étendrons à une nouvelle mesure hybride combinant la distance de Rada [2] (la distance de deux concepts issus de l'ontologie) avec la quantité d'information partagée entre les deux concepts dans le corpus des données (plus la quantité d'information partagée est importante plus les concepts sont similaires). On notera l'équation suivante :

$$Sim(c_i, c_j) = \frac{2*[E(c_i) \cap E(c_j)]}{E(c_i) + E(c_j)} * \frac{1}{1 + dist(c_i, c_j)} \quad (1)$$

$Sim(c_i, c_j)$ la valeur de similarité entre le concept $c_i$ et le concept $c_j$

$E(c_i) \cap E(c_j)$ les plans de vidéo indexées par $c_i$ et $c_j$

$E(c_i) + E(c_j)$ la somme de plans de vidéos indexées par $c_i$ et les plans de vidéos indexées par $c_j$

$dist(c_i, c_j)$ : la distance entre $c_i$ et $c_j$ = nb arcs séparant les deux concepts dans l'ontologie

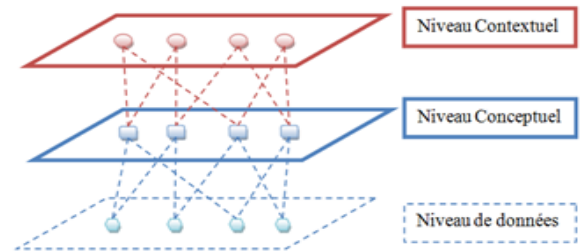

Figure 2. Fichier XML représentant les différents contextes

### 2.2. Organisation de la base de données

*) Structuration de données selon trois niveaux (niveau contextuel, niveau conceptuel, niveau de données brutes).
*) Pondération de concepts en utilisant une formule bien déterminée.[1]

Figure 3. Structuration de nos données

On utilise l'approche basée sur TF*IDF car cette dernière demeurant la plus employée par les systèmes de recherche d'information.

$tf(c_i, v_j) = P_l(c_i, v_j)$ Une pondération globale s'intéresse aux informations concernant les concepts et dépend de la collection de documents (2) $f(c_i, v_j) = P_2(c_i, v_j)$ : une pondération locale permet de mesurer la représentativité locale d'un concept. (3)

$$P(c_i, v_j) = P_1 * P_2$$
$$P(c_i, v_j) = \left[\frac{nombre_{shots(c_i,v_j)}}{n}\right] * \left[\frac{nombre_{concepts_{simi(c_i)}}}{nombre_{concepts(v_j)} * N}\right]$$
$$(4 = 1 * 2)$$

$n$: Nombre total de plans dans la vidéo $V_j$
$N$: Nombre de vidéos contenants le concept $C_i$

**2.3. Techniques de visualisation interactives**

**2.3.1. Utilisation de technique de nuage de mots clés**

Une fois la base est structurée, il est nécessaire de représenter les informations obtenues de façon à les rendre facilement accessibles par l'utilisateur. L'interface est donc la finalisation de la conception d'un système de navigation. Elle doit permettre `a l'utilisateur de visualiser le maximum d'informations apportées par les étapes précédentes, d'intuitivement de comprendre les mécanismes mis en œuvre par le système de la navigation. Nous devons tout d'abord représenter une vue d'ensemble de la base audio-visuelle pour faciliter sa compréhension globale. Dans ce contexte, nous proposons d'utiliser une technique de nuage de mots clés (Tag Cloud en anglais) qui fournit une représentation visuelle des concepts les plus pertinents (déjà calculés en phase d'organisation de données) permet d'apporter un résumé sémantique du corpus, dans lequel les concepts clefs évoqués s'affichent avec de taille de police plus grandes indépendamment de sa fréquence. Une fois l'utilisateur sélectionne le concept désiré peut se déplacer au deuxième niveau de la navigation en affichant les concepts similaires. Cette technique évite non seulement la surcharge de l'écran de la navigation, mais elle permet également la désorientation de l'utilisateur.

**2.3.2. Similarité sémantique vers la similarité spéciale**

Suivant l'esprit de la visualisation (paradigme de Schneiderman), nous proposons une représentation visuelle 2D permet de traduire la similarité sémantique inter-vidéos en similarité spéciale. Pour fournir une guide dont l'objectif est d'aider l'utilisateur pendant sa tâche d'exploration du corpus (couche inférieure de note système de navigation). Dans le cadre de cette approche, l'utilisateur est assisté par une articulation judicieuse entre la vue globale et la vue locale maintenant sa carte mentale et par la suggestion de parcours cohérents à travers les documents audio-visuels.

**3. Système gestuel pour les personnes handicapées**

Nous avons donc décidé de créer un système de recherche, offrant en plus des commandes usuelles, différents types d'accès au système, en fonction du handicape de la personne. En effet, l'application se compose d'un système de recherche, mais offre l'accès aux commandes par le biais de la voix ou les gestes. De cette manière, si une personne a un handicap moteur, elle peut accéder aux services de notre système par sa voix ou, il peut utiliser les gestes pour exprimer son besoin. Le but de ce système est donc de s'adapter à tout type du handicap. La figure montre les éléments constitutifs de notre application. Bien évidemment nous voyons le capteur de gestes **Kinect** comme élément central. L'ordinateur est connecté au notre système exécutant les actions demandées par l'utilisateur.

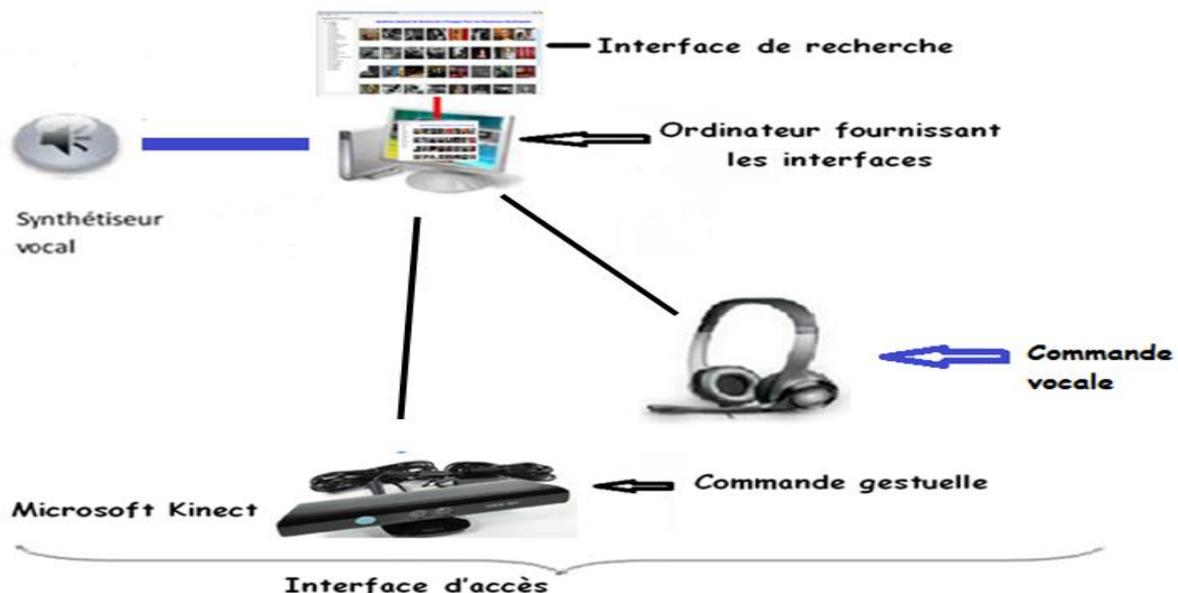

Figure 4. Système gestuel pour la recherche de vidéo

**4. Interface du système :** L'interface illustrée par la figure 5 représente l'interface de menu principal de notre système de recherche d'image.

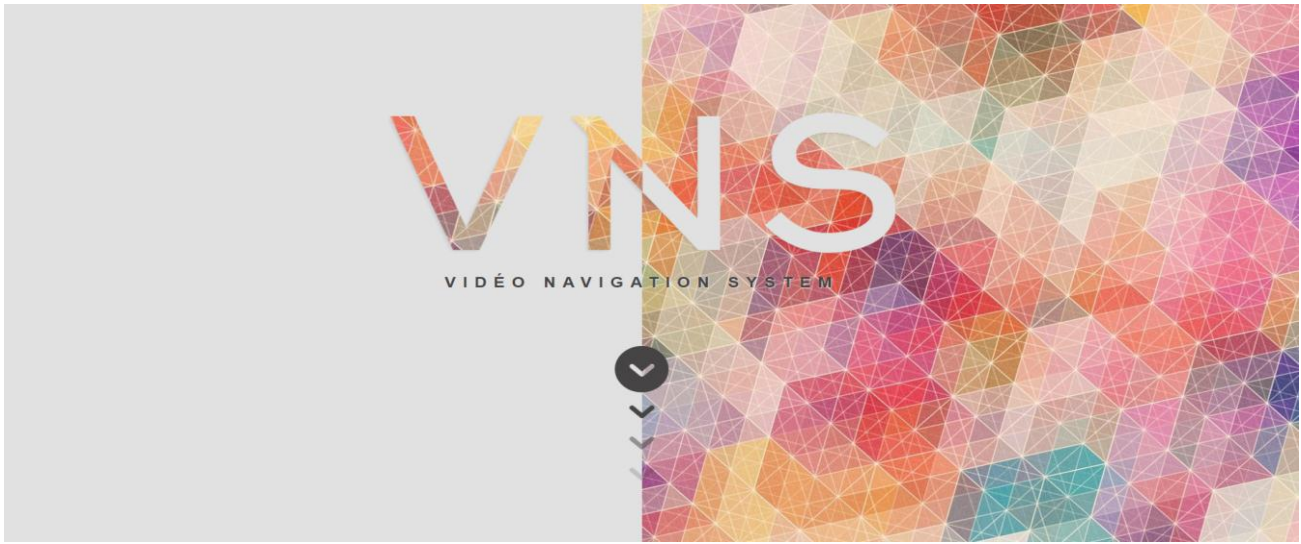

Figure 5. Menu principal

L'interface de navigation est composée de 3 parties :( A) pour l'affichage des contextes, une fois l'utilisateur a choisi le contexte que l'intéresse, un ensemble de concepts similaires est affiché au partie(B). La partie (C) est consacrée à l'affichage de vidéos relatives au concept choisi, ces vidéos relatives sont ordonnées selon le degré d'appartenance de concept dans les vidéos-mêmes.

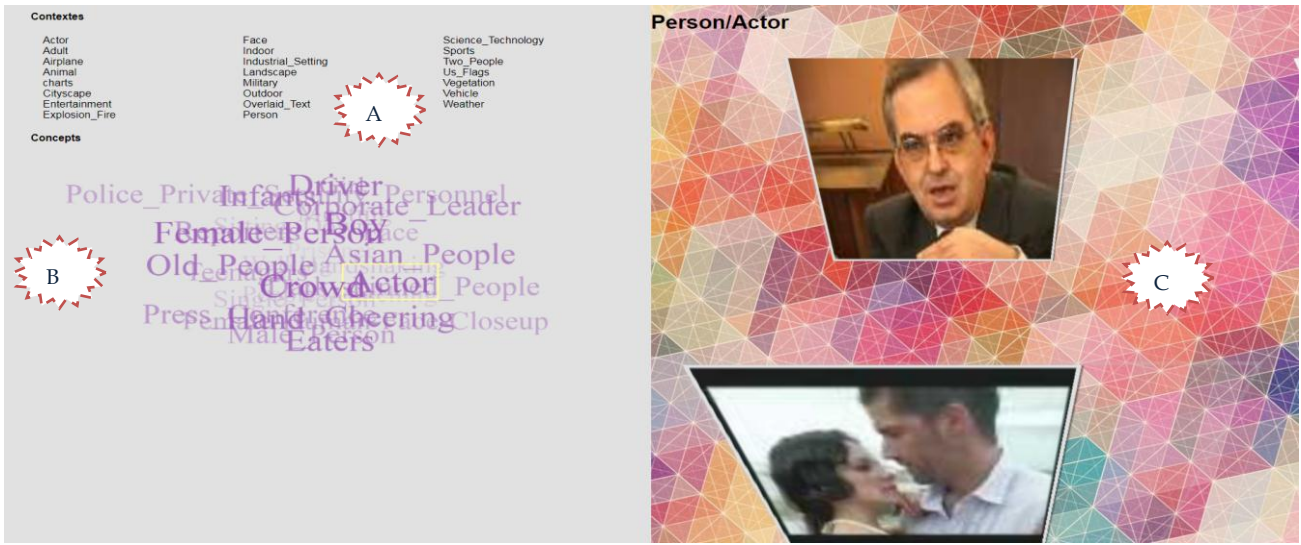

Figure 6. Recherche par navigation

**5. Evaluation**

La méthode la plus courante pour évaluer une interface homme machine est la réalisation d'une étude utilisateur. Cette technique consiste à proposer un questionnaire aux utilisateurs afin qu'ils répondent à chaque question généralement par une note. Ce questionnaire est inspiré des propositions de Schneiderman [3] en matière d'évaluation d'interfaces utilisateur. Ce principe a été notamment adopté dans [4] et puisque notre système ne possède pas une phase de formulations de requête, nous pensons que la méthode la plus adéquaté pour évaluer le système proposé est de réaliser un questionnaire.

Les questions ont été choisies de telle façon à garder un test assez court et peu contraignant pour l'utilisateur. En effet, nous souhaitons qu'il puisse répondre à ce test en environ de 10 minutes. Il est note que chaque question possède une note varie entre 1et 5 vis-à-vis le point de vue de l'évaluateur. Le questionnaire ainsi réalisé et proposé aux utilisateurs est présenté dans le tableau 1 .15 utilisateurs ont participé à un ce test : quatre parmi eux sont des membres de l'équipe de recherche au laboratoire de SETIT à Sfax et le reste sont des étudiants.

Tout d'abord, les réponses des utilisateurs aux questions relatives au système sont synthétisées dans le tableau 2. Ensuite, nous proposons une représentation graphique (sous la forme d'histogramme) de ces réponses. Une interprétation des résultats de ce test est présentée à la fin de cette section. Nous rappelons que le note « 5 » correspond à la meilleure réponse, cependant, le note « 1 » présente la mauvaise réponse.

| Questions | | 1 | 2 | 3 | 4 | 5 | |
|---|---|---|---|---|---|---|---|
| 1-Comment trouver vous ce système ? | Frustrant | | | 4 | 6 | 5 | Satisfaisant |
| 2-Utilisation de ce système | Difficile | | | 2 | 9 | 4 | Facile |
| 3-Vitesse de système | Trop lent | | | 5 | 7 | 3 | Assez rapide |
| 4-Comment trouver ce système ? | Confus | | | 6 | 8 | 1 | Clair |
| 5-Fiabilité de système | peu fiable | | | 3 | 8 | 4 | Fiable |
| 6-Le technique nuage de concepts simplifie la navigation | Pas de tous | | 1 | 3 | 5 | 6 | Beaucoup |
| 7-La similarité inter-concepts | Confus | | 1 | 5 | 8 | 1 | Clair |
| 8-La représentation 2D de similarité inter-vidéos. | Inadaptée | | 1 | 3 | 4 | 7 | Adaptée |
| 9-Compréhension de similarité inter-vidéos | difficile | | 2 | 5 | 7 | 1 | Facile |
| 10-Organisation de l'information dans l'interface | Confus | | 3 | 3 | 7 | 2 | très clair |
| 11-Navigation | difficile | | 1 | 2 | 7 | 5 | Facile |
| 12- Trouvez-vous les concepts pertinentes utiles pour se repérer ? | Pas de tout | 2 | | 4 | 7 | 2 | Enormément |
| 13-Comment trouvez la technique de zoom | Inutile | 2 | 1 | 4 | 4 | 4 | Utile |
| 14- le technique de zoom présente plus de détails sur le contenu de vidéo | Pas de tout | 2 | | 4 | 6 | 3 | Beaucoup |
| 15- Que pensez-vous à l'ajout d'historique | Inutile | 3 | 3 | 2 | 4 | 3 | Utile |
| 16- Que pensez-vous à l'utilisation de graphe pour visualiser la similarité inter-vidéos au lieu de sphère 3D | Inutile | 1 | 3 | | 4 | 7 | Utile |
| 17- Que pensez-vous à l'ajout d'aide | Inutile | 1 | 1 | 3 | 2 | 8 | Utile |
| 18-Apprendre à utiliser le système | Difficile | | 1 | 3 | 9 | 2 | Facile |
| 19-Destiné à tous les utilisateurs | Jamais | 1 | 2 | 5 | 3 | 4 | Toujours |

Tableau 2 : Résultats de test d'évaluation

| Note | 1 | 2 | 3 | 4 | 5 |
|---|---|---|---|---|---|
| **Nombre total de vote** | 9 | 20 | 66 | 115 | 72 |
| **Moyenne** | 3% | 7% | 23% | 41% | 25% |

Tableau 3 **:** Distribution des votes sur les 19 questions.

La moyenne de note est calculée par la formule suivante :

$$Moyenne(note\ i) = \frac{Nombre\ des\ votes\ correspond\ au\ note\ i}{Nombre\ total\ des\ votes} * 100$$

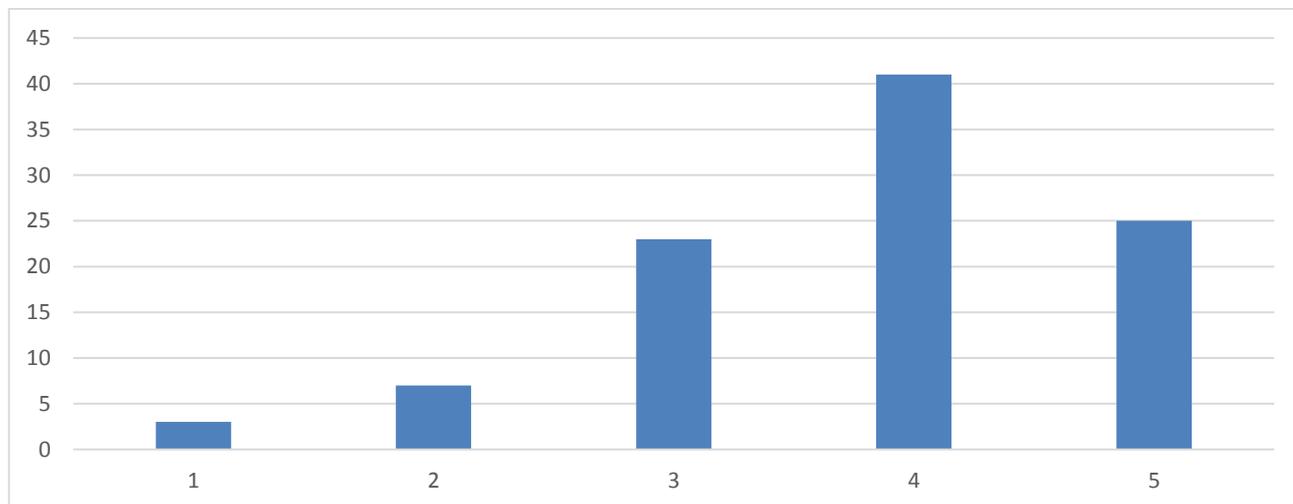

Figure 7. Résultats de test de l'évaluation

L'histogramme présenté dans la figure 7 montre que les maximums de votes sont focalisés dans le notes centrale (4). La conclusion qu'nous pouvons la tirer à partir les résultats de ce test est que les utilisateurs ne trouvent pas des difficultés à utiliser notre système de navigation. Néanmoins, il est nécessaire d'améliorer certains points tels que la clarté de similarité inter-concepts par l'utilisation des graphes conceptuels et l'ajout d'un historique permet de conserver le chemin de la navigation.

## 6. Conclusion

Dans cet article, nous avons validé notre proposition système de recherche vidéo à partir d'un corpus de documents audiovisuels (TRECVID 2010) caractérisé par sa taille et l'importance de son contenu hétérogénéité. Nous avons développé un système de recherche de vidéos d'accéder facilement à la vidéo désirée. Ainsi, le système fournit un système gestuel pour les personnes handicapées.